\documentclass[pdflatex,sn-mathphys-num]{sn-jnl}

\usepackage{graphicx}
\usepackage{multirow}
\usepackage{amsmath,amssymb,amsfonts}
\usepackage{mathrsfs}
\usepackage[title]{appendix}
\usepackage{xcolor}
\usepackage{textcomp}
\usepackage{manyfoot}
\usepackage{booktabs}
\usepackage{algorithm}
\usepackage{algorithmicx}
\usepackage{algpseudocode}
\usepackage{listings}
\usepackage{caption}
\usepackage{bm} 

\begin{document}


\title[SCAR: Semantic Cardiac Adversarial Representation]{SCAR: Semantic Cardiac Adversarial Representation via Spatiotemporal Manifold Optimization in ECG}

\author[1]{\fnm{Shunbo} \sur{Jia}}\email{2240003657@student.must.edu.mo}
\author*[2]{\fnm{Caizhi} \sur{Liao}}\email{liaocaizhi@suat-sz.edu.cn}

\affil[1]{\orgdiv{Faculty of Innovation Engineering}, \orgname{Macau University of Science and Technology}, \city{Macau}, \country{China}}
\affil[2]{\orgdiv{School of Biomedical Engineering}, \orgname{Shenzhen University of Advanced Technology}, \city{Shenzhen}, \country{China}}

\abstract{
Deep learning models for Electrocardiogram (ECG) analysis have achieved expert-level performance but remain vulnerable to adversarial attacks. However, applying Universal Adversarial Perturbations (UAP) to ECG signals presents a unique challenge: standard imperceptible noise constraints (e.g., $10\mu V$) fail to generate effective universal attacks due to the high inter-subject variability of cardiac waveforms. Furthermore, traditional "invisible" attacks are easily dismissed by clinicians as technical artifacts, failing to compromise the human-in-the-loop diagnostic pipeline. In this study, we propose \textbf{SCAR (Semantic Cardiac Adversarial Representation)}, a novel UAP framework tailored to bypass the clinical "Human Firewall." Unlike traditional approaches, SCAR integrates spatiotemporal smoothing ($W=25$, approx. 50ms), spectral consistency ($<15$ Hz), and anatomical amplitude constraints ($<0.2$mV) directly into the gradient optimization manifold.
\textbf{Results:} We benchmarked SCAR against a rigorous baseline (Standard Universal DeepFool with post-hoc physiological filtering). While the baseline suffers a performance collapse ($\sim$16\% success rate on transfer tasks), \textbf{SCAR} maintains robust transferability (\textbf{58.09\%} on ResNet) and achieves \textbf{82.46\%} success on the source model. Crucially, clinical analysis reveals an \textbf{emergent targeted behavior}: SCAR specifically converges to forging Myocardial Infarction features (\textbf{90.2\%} misdiagnosis) by mathematically reconstructing pathological ST-segment elevations. Finally, we demonstrate that SCAR serves a dual purpose: it not only functions as a robust data augmentation strategy for \textbf{Hybrid Adversarial Training}, offering optimal clinical defense, but also provides \textbf{effective educational samples} for training clinicians to recognize low-cost, AI-targeted semantic forgeries.
}

\keywords{ECG, Universal Adversarial Perturbation, Semantic Morphological Forgery, Manifold Optimization, Adversarial Defense}

\maketitle


\section{Introduction}\label{sec:intro}
Cardiovascular diseases remain the leading cause of death globally \cite{GBD2023}. While Deep Neural Networks (DNNs) have demonstrated remarkable accuracy in arrhythmia detection \cite{chen2024coordinated}, their robustness against adversarial examples remains a critical safety concern \cite{szegedy2013intriguing, goodfellow2014explaining}.

Current adversarial research in ECG largely follows the image domain paradigm, prioritizing imperceptibility by restricting perturbations to microscopic levels (e.g., $L_\infty \le 10\mu V$) \cite{han2020deep}. While effective for purely algorithmic attacks, we argue that this constraint renders \textbf{Universal Adversarial Perturbations (UAP)} \cite{moosavi2017universal, huang2024texture} clinically irrelevant. In real-world workflows, the cardiologist acts as a \textbf{"Human Firewall."} Traditional high-frequency adversarial noise appears as non-biological "jitter" to experts and is easily filtered out. To fundamentally assess the safety of AI diagnostics, we must consider attacks that penetrate this human layer by exhibiting biological plausibility.

In this paper, we challenge this paradigm with \textbf{SCAR}. We name our method \textbf{SCAR} (\textbf{S}emantic \textbf{C}ardiac \textbf{A}dversarial \textbf{R}epresentation) to reflect its dual nature. Academically, it denotes the generation of adversarial examples that possess valid semantic meaning in the cardiac domain. Clinically, the term alludes to \textit{"Myocardial Scarring"}—the permanent structural damage following a heart attack. Just as physical scarring distorts the heart's electrical conduction path, our algorithm mathematically "scars" the healthy ECG waveform with pathological deformations. 

Rather than viewing this solely as a security threat, we propose SCAR as a constructive tool for \textbf{systemic robustness enhancement}. By generating semantic morphological forgeries, SCAR acts as a rigorous "stress test" or an advanced form of adversarial data augmentation. This approach exposes shared vulnerabilities between AI feature extraction and human visual diagnosis, enabling the development of models that are robust not just to digital noise, but to deceptive semantic manipulations.

To the best of our knowledge, SCAR is the \textbf{first} UAP framework specifically designed to generate \textbf{semantic morphological forgeries} via \textbf{in-process gradient manifold projection}, explicitly targeting the shared vulnerability between AI models and human clinical interpretation.

\section{Methodology}\label{sec:method}

\subsection{Problem Formulation}
Let $\mathcal{D} = \{(\mathbf{x}_i, y_i)\}_{i=1}^N$ denote the dataset of ECG recordings, where $\mathbf{x} \in \mathbb{R}^{L \times T}$ represents the $L$-lead ECG signal with length $T$, and $y$ is the ground truth label. Let $f_\theta: \mathbb{R}^{L \times T} \rightarrow \mathbb{R}^K$ be a deep neural network classifier parameterized by $\theta$.

Unlike traditional adversarial attacks that optimize a unique perturbation for each input, we seek a \textbf{Universal Adversarial Perturbation (UAP)}, denoted as $\mathbf{v}$, that deceives the model across the entire data distribution. The objective function is defined as:

\begin{equation}
    \max_{\mathbf{v}} \underset{(\mathbf{x}, y) \sim \mathcal{D}}{\mathbb{E}} [\mathcal{L}(f_\theta(\mathbf{x} + \mathbf{v}), y)] \quad \text{s.t.} \quad \mathbf{v} \in \mathcal{S}_{phy}
\end{equation}

where $\mathcal{L}$ is the cross-entropy loss, and $\mathcal{S}_{phy}$ represents the set of physiologically plausible constraints.

\subsection{Baseline Definition: Standard UAP with Post-hoc Constraints}
To rigorously evaluate the necessity of our proposed manifold optimization, we compare SCAR against the standard Universal Adversarial Perturbation algorithm \cite{moosavi2017universal}, which aggregates perturbations generated by \textbf{DeepFool} \cite{moosavi2016deepfool}. 

Crucially, to create a fair "Strong Baseline," we do \textbf{not} restrict this baseline to the standard imperceptible range ($10\mu V$). Instead, we allow the baseline to utilize the \textbf{same amplitude magnitude} (0.1--0.2 mV) as SCAR. The fundamental difference lies in the optimization strategy: the Baseline employs a \textbf{"Generate-then-Filter" (Post-hoc)} paradigm. It calculates the standard UAP vector $\mathbf{v}_{raw}$ and then applies physiological constraints $\Phi$ (low-pass filtering and smoothing) as a post-processing step:
\begin{equation}
    \mathbf{v}_{base} = \Phi(\mathbf{v}_{raw}) = \text{Clip}_{amp}(\mathcal{F}_{low}(\mathbf{v}_{raw}))
\end{equation}
This baseline represents the best possible performance achievable by applying traditional geometric attacks to the ECG domain without modifying the core optimization objective.

\subsection{Proposed Method: SCAR}
SCAR diverges from the baseline by integrating physiological constraints directly into the optimization process via \textbf{In-process Gradient Manifold Projection}.

\subsubsection{Physiological Constraints Set}
We define the feasible set $\mathcal{S}_{phy}$ through three distinct constraints derived from clinical sensitivity analysis:
\begin{enumerate}
    \item \textbf{Anatomical Amplitude Constraint ($\mathcal{C}_{amp}$):} We define a lead-wise constraint vector $\boldsymbol{\epsilon} \in \mathbb{R}^{L}$. Based on our experiments (Sec. \ref{sec:sensitivity}), we observe that the V1 chest lead exhibits lower amplitudes comparable to limb leads. Therefore, we set $\epsilon_i = 0.1$ mV for \textbf{Limb Leads and V1}, and $\epsilon_i = 0.2$ mV for the remaining \textbf{Chest Leads (V2--V6)}.
    \item \textbf{Spectral Consistency ($\mathcal{C}_{freq}$):} The perturbation energy must be confined to the physiological bandwidth $[0, 15]$ Hz.
    \item \textbf{Morphological Smoothness ($\mathcal{C}_{morph}$):} To mimic cardiac depolarization dynamics \cite{golany2021ecg}, we enforce temporal continuity via a smoothing window $W=25$ (50ms).
\end{enumerate}

\subsubsection{In-Process Gradient Smoothing (Sobolev Gradient)}
We employ Normalized Stochastic Gradient Descent (SGD) to update the universal perturbation $\mathbf{v}$. Let $\mathbf{g}_t = \nabla_{\mathbf{v}} \mathcal{L}$ be the raw gradient at step $t$. Instead of updating $\mathbf{v}$ directly with this $L_2$ gradient, which often contains high-frequency noise, we project the gradient onto the smooth signal manifold.

\textbf{Theoretical Foundation:} Mathematically, this smoothing operation corresponds to optimizing the perturbation within the \textbf{Sobolev space} rather than the standard Euclidean space. As proven by Neuberger et al. \cite{neuberger2010sobolev}, the Sobolev gradient $\nabla_{S} \mathcal{L}$ is related to the standard gradient $\nabla_{L_2} \mathcal{L}$ via convolution with a smoothing kernel $K$:
\begin{equation}
    \nabla_{S} \mathcal{L} = K \ast \nabla_{L_2} \mathcal{L}
\end{equation}
While gradient smoothing techniques have been utilized in computer vision to improve translation invariance \cite{dong2019evading} or physical realizability \cite{brown2017adversarial}, SCAR fundamentally repurposes this mechanism for the cardiac domain. Instead of seeking spatial robustness, we employ this operator to strictly confine the optimization trajectory within the manifold of \textbf{physiologically continuous} temporal dynamics.

In our implementation, we define the kernel operator as a composite of spectral filtering and temporal smoothing:
\begin{equation}
    \mathbf{g}'_t = \mathcal{K}_{temp} \ast (\mathcal{F}_{low} (\mathbf{g}_t))
\end{equation}

where $\mathcal{F}_{low}$ is a Butterworth low-pass filter, and $\mathcal{K}_{temp}$ is a temporal Gaussian smoothing kernel implemented via convolution ($\ast$). The update rule is:

\begin{equation}
    \mathbf{v}_{t+1} = \Pi_{\boldsymbol{\epsilon}} \left( \mathbf{v}_t + \alpha \cdot \frac{\mathbf{g}'_t}{\|\mathbf{g}'_t\|_2} \right)
\end{equation}

where $\Pi_{\boldsymbol{\epsilon}}$ is the element-wise projection onto the anatomical amplitude constraints. 
\textbf{Mechanism of Universality:} Unlike instance-specific attacks (e.g., PGD) that reset the perturbation for each input, SCAR accumulates gradient information ($\mathbf{g}'_t$) across the entire data distribution. This allows $\mathbf{v}$ to learn global pathological features that exploit shared vulnerabilities in the model, rather than overfitting to the noise tolerance of individual samples.

\section{Experiments and Results}\label{sec:exp}

\subsection{Experimental Setup}
We evaluated our method on three datasets: \textbf{PTB-XL} \cite{wagner2020ptb} (Source), \textbf{CPSC2018} \cite{liu2018open}, and the \textbf{Chapman-Shaoxing 12-Lead ECG Database} \cite{zheng202012}. 
\textbf{Baseline Configuration:} We compare SCAR against the aforementioned Universal DeepFool Baseline. Both methods utilize identical constraint parameters ($W=25, F_c=15$Hz) to isolate the impact of the optimization strategy.

\subsection{Parameter Sensitivity and Optimization}\label{sec:sensitivity}
Selecting appropriate amplitude constraints is crucial for balancing attack efficacy and signal fidelity. We conducted a sensitivity analysis by varying the amplitude limit for \textbf{chest leads V2--V6} from $0.02$ mV to $0.60$ mV (while keeping limb leads and V1 fixed at 0.1 mV).

\begin{figure}[t!]
    \centering
    \includegraphics[width=0.95\textwidth]{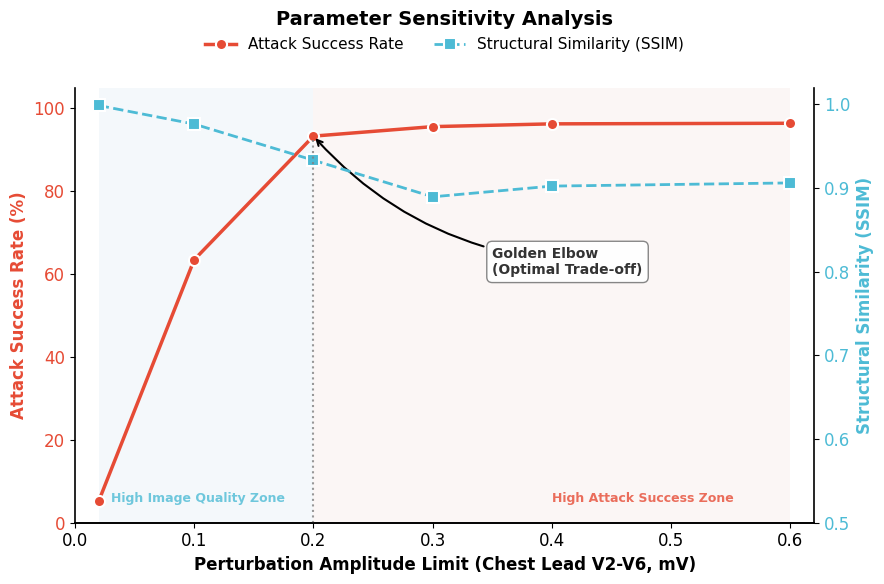}
    \caption{\textbf{Parameter Sensitivity Analysis.} The dual-axis plot reveals the trade-off between Attack Success Rate (Red) and Structural Similarity (Blue) as the V2--V6 amplitude limit varies. A "Golden Elbow" is observed at 0.2 mV, representing the optimal equilibrium validated by both statistical metrics and clinical assessment.}
    \label{fig:sensitivity}
\end{figure}

As illustrated in \textbf{Figure \ref{fig:sensitivity}}, we identified a "Golden Elbow" at \textbf{0.2 mV}. At this operating point, the Attack Success Rate (red line) surges to over 90\%, while the Structural Similarity Index (SSIM, blue dashed line) remains high ($>0.93$).

\textbf{Clinical Justification:} Crucially, this threshold (0.2 mV) is not merely a statistical optimum but is grounded in clinical validity. Clinical assessment by cardiologists confirmed that perturbations below this threshold in chest leads remain consistent with plausible pathological morphology (specifically ST-elevation) without introducing visible non-biological artifacts. This ensures the generated samples pass the "Visual Turing Test," maintaining the dual-attack capability against both algorithms and human experts.

\subsection{Main Result: Attack Effectiveness}
Table \ref{tab:comparison} presents the attack success rates. The Baseline (Universal DeepFool + Post-hoc Filter) proves to be a rigorous competitor: due to physiological filtering, it acquires semantic properties that allow it to slightly outperform SCAR in transferability to the Conv-ViT architecture (34.78\% vs. 30.95\%). However, this "post-hoc" semantic alignment is fragile; the Baseline struggles significantly against the deep residual features of ResNet, dropping to $\sim$16\% success rate. In contrast, SCAR demonstrates superior stability across all architectures, achieving \textbf{58.09\%} transferability on ResNet, validating that manifold-aware optimization yields more generalizable morphological features than geometric projection alone.

\begin{table}[t!]
\centering
\caption{\textbf{Comparison of Attack Success Rates (Fooling Rate, \%).} Comparisons are made against a physiologically constrained Universal DeepFool baseline. SCAR demonstrates superior transferability across architectures. \textit{Note: "Conv-ViT" refers to a hybrid architecture utilizing convolutional embedding layers for local morphological feature extraction followed by a transformer encoder for global context.}}\label{tab:comparison}
\begin{tabular}{@{}llccc@{}}
\toprule
\textbf{Target Dataset} & \textbf{Model Arch.} & \textbf{Baseline (Filtered DeepFool)} & \textbf{SCAR (Ours)} & \textbf{Improvement} \\
\midrule
PTB-XL (Source) & CNN & 80.27 & \textbf{82.46} & +2.19 \\
\textbf{PTB-XL (Transfer)} & \textbf{ResNet} & 16.60 & \textbf{58.09} & \textbf{+41.49} \\
PTB-XL (Cross-Arch) & Conv-ViT & \textbf{34.78} & 30.95 & -3.83 \\
CPSC2018 & CNN & 25.26 & \textbf{29.20} & +3.94 \\
Chapman-Shaoxing & CNN & 31.12 & \textbf{35.38} & +4.26 \\
\botrule
\end{tabular}
\end{table}

\section{Discussion and Mechanism Analysis}\label{sec:discussion}

\subsection{Mechanism: Bypassing the Human Firewall via Semantic Rewriting}
Traditional adversarial attacks rely on "invisible noise" paradigms that, while mathematically effective against CNNs, fail in clinical practice. Such high-frequency artifacts are easily identified and discarded by clinicians—the "Human Firewall."

\begin{figure}[t!]
    \centering
    \includegraphics[width=\linewidth]{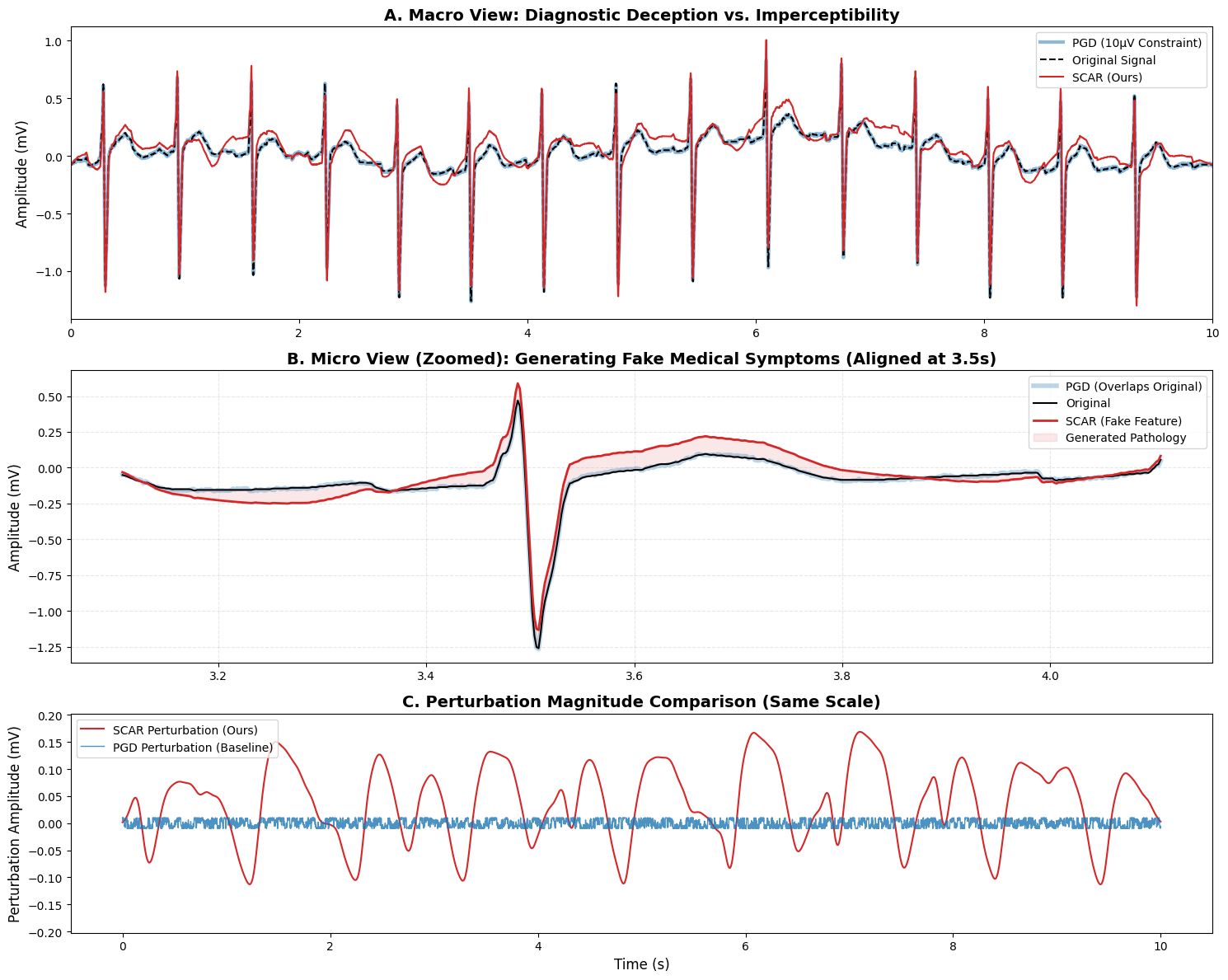}
    \caption{\textbf{Qualitative Mechanism Analysis: Semantic Rewriting vs. Traditional Noise.} 
    \textbf{A. Macro View:} The traditional PGD attack (blue reference) is invisible, whereas SCAR (red) visibly modifies the waveform morphology.
    \textbf{B. Micro View (Zoomed):} SCAR generates a "fake" ST-segment elevation (Pathological Mimicry), whereas the traditional attack appears as random jitter.
    \textbf{C. Perturbation Magnitude:} SCAR utilizes large-scale, smooth perturbations within the physiological manifold, distinct from the microscopic noise of traditional methods.}
    \label{fig:mechanism}
\end{figure}

SCAR fundamentally shifts this paradigm towards \textbf{Semantic Morphological Forgery}. As visualized in \textbf{Figure \ref{fig:mechanism}}, SCAR does not merely add noise; it rewrites the signal's semantics.
\textit{Note: While our performance baseline is DeepFool, Figure \ref{fig:mechanism} utilizes a standard PGD attack ($L_\infty \le 10\mu V$) as a visual reference to represent the "Invisible Noise" paradigm.}

\subsection{Spectral Consistency and Targeted Behavior}
\begin{figure}[t!]
    \centering
    \includegraphics[width=0.85\textwidth]{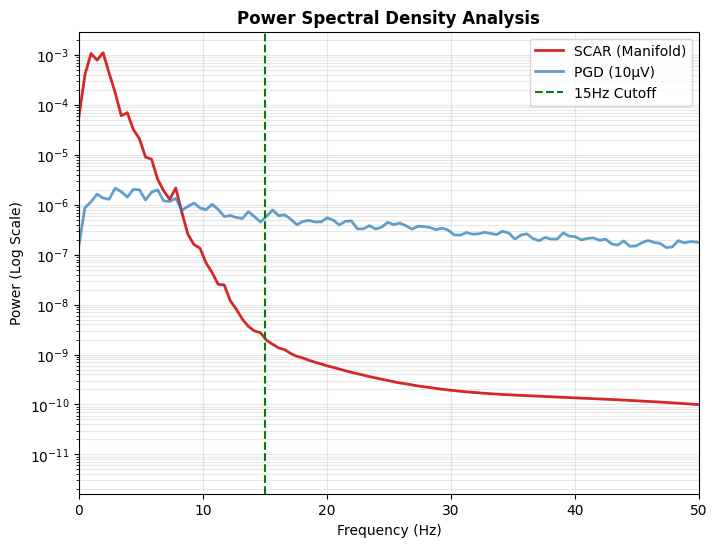}
    \caption{\textbf{Power Spectral Density (PSD) Analysis.} 
    The SCAR perturbation (red) respects the 15Hz physiological bandwidth constraint (green line). In contrast, the traditional attack (blue) exhibits high-frequency noise characteristics.}
    \label{fig:spectrum}
\end{figure}

A key requirement for physically realizable adversarial attacks is spectral consistency. As shown in \textbf{Figure \ref{fig:spectrum}}, the SCAR perturbation (red line) concentrates its energy strictly below the 15Hz cutoff.

\begin{figure}[t!]
    \centering
    \includegraphics[width=0.95\textwidth]{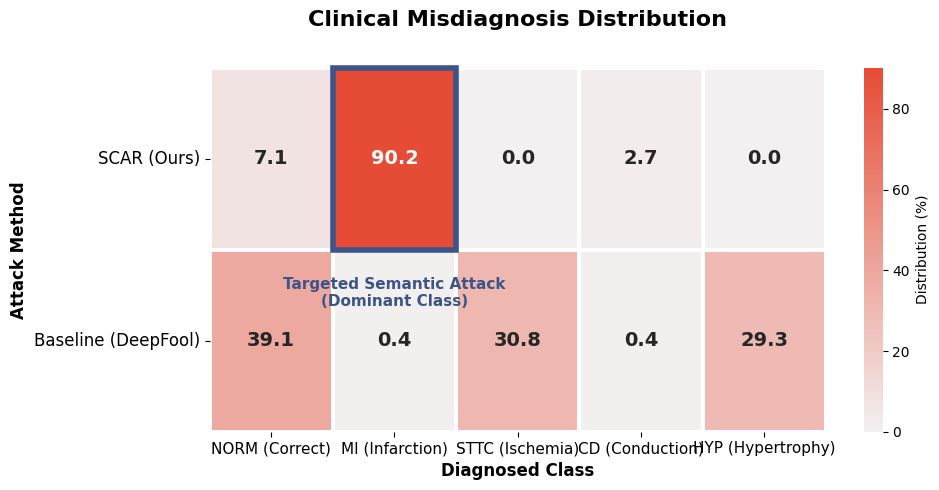}
    \caption{\textbf{Clinical Misdiagnosis Distribution.} 
    The heatmap illustrates the class distribution of successful attacks. \textbf{SCAR (Top Row)} exhibits an \textit{emergent targeted behavior}, forcing \textbf{90.2\%} of samples into the Myocardial Infarction (MI) class. The \textbf{Baseline (Bottom Row)} results in a dispersed distribution.}
    \label{fig:misdiagnosis}
\end{figure}

Furthermore, \textbf{Figure \ref{fig:misdiagnosis}} reveals a critical insight into the \textbf{Biological-Digital Dual Vulnerability}. Although formulated as \textbf{untargeted} attacks, SCAR exhibits an \textbf{emergent targeted behavior}, with \textbf{90.2\%} of successful attacks converging to the \textbf{Myocardial Infarction (MI)} class. We attribute this to the manifold constraints: generating ST-elevation represents the optimization "path of least resistance" on the loss landscape. This implies that AI models, like human doctors, heavily rely on prominent local morphological features rather than global context, making them susceptible to this form of semantic forgery.

\subsection{Controllability of Perceptibility}
A notable advantage of the SCAR framework is the controllability of the generated perturbation's perceptibility via spectral hyperparameters. While our primary configuration uses a 15Hz cutoff to achieve high-impact semantic rewriting, the framework allows for stricter constraints.
We experimentally observed that by tightening the spectral constraint (e.g., reducing the cutoff to 5Hz), SCAR generates UAPs that are visually indistinguishable from the original signal. Crucially, unlike PGD-based methods \cite{madry2018towards} which often introduce high-frequency "square-wave" artifacts, SCAR's manifold projection ensures smoothness. This demonstrates that SCAR is a generalized framework capable of spanning the spectrum from invisible noise to visible semantic rewriting.

\subsection{Data Quality and Implicit Adversarial Robustness}
An intriguing phenomenon was observed in our cross-dataset evaluation: SCAR generated from the Chapman-Shaoxing dataset exhibited significantly lower transferability (e.g., 8.41\% FR on CPSC) compared to attacks from high-quality sources like PTB-XL. We attribute this to the inherent signal quality of the Chapman-Shaoxing dataset, which contains substantial high-frequency noise. Training on such noisy data forces the model to ignore local fluctuations and focus on robust, global morphological features, effectively functioning as a form of \textbf{"Implicit Adversarial Training."}

This observation aligns perfectly with our defense experiments (Section \ref{sec:defense}), where explicit PGD Adversarial Training provided strong defense against SCAR. Both scenarios—training on naturally noisy data (Chapman-Shaoxing) and training on artificially perturbed data (PGD-AT)—smooth the decision boundary, making the model desensitized to the subtle manifold manipulations introduced by SCAR. This suggests that while high-quality data (PTB-XL) allows SCAR to learn highly transferable semantic features, noisy data inadvertently fortifies models against such attacks, albeit at the cost of generating less effective adversarial examples.

\subsection{Ethical Considerations}
The development of algorithms capable of generating physiologically plausible ECG forgeries introduces potential dual-use risks. To mitigate misuse, this research was conducted under strict ethical guidelines. The primary objective is "Red Teaming"—identifying vulnerabilities in diagnostic AI systems to enable the development of more robust models. All code and generated samples will be released solely to credentialed academic and clinical researchers under a responsible disclosure agreement. Furthermore, we advocate for the inclusion of semantic adversarial examples in the regulatory approval process for medical AI software (SaMD) to ensure safety against both natural anomalies and malicious manipulations.

\section{Defense and Mitigation}\label{sec:defense}

\begin{figure}[t!]
    \centering
    \includegraphics[width=\textwidth]{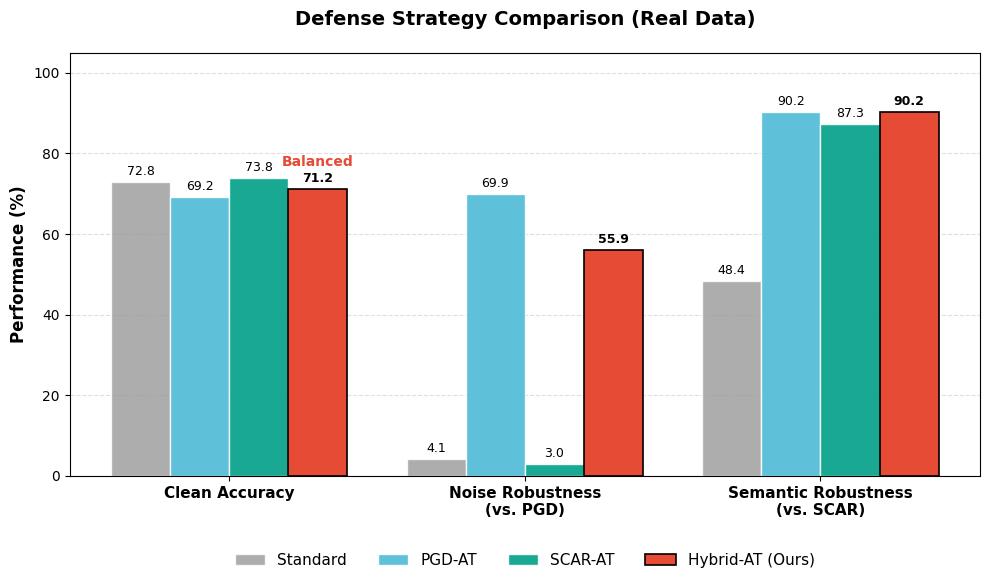}
    \caption{\textbf{Defense Strategy Comparison.} 
    A grouped bar chart illustrating the trade-offs between clean accuracy and robustness. \textbf{Hybrid-AT} (red) is highlighted as the balanced choice, offering comprehensive protection with higher clean accuracy.}
    \label{fig:defense}
\end{figure}

Given the vulnerability of standard models (4.1\% robustness against PGD), we evaluated three Adversarial Training (AT) strategies. The results in \textbf{Figure \ref{fig:defense}} reveal a striking \textbf{asymmetry in robustness generalization}.

\begin{itemize}
    \item \textbf{Asymmetric Generalization:} \textbf{PGD-AT} (blue) \cite{madry2018towards} generalizes well to semantic attacks (90.2\% robustness against SCAR). However, the reverse is not true: \textbf{SCAR-AT} (green) completely fails against noise attacks (2.98\%).
    \item \textbf{Optimal Equilibrium (Hybrid-AT):} Our proposed \textbf{Hybrid-AT} (red) achieves the most clinically viable equilibrium, recovering significant Clean Accuracy (71.2\%) while maintaining state-of-the-art defense (90.2\%) \cite{shafahi2020universal}. By incorporating SCAR samples, the model effectively learns to distinguish between true pathology and semantic forgery.
\end{itemize}

\section{Conclusion}\label{sec:conc}
We presented SCAR, a method that shifts the adversarial paradigm from "invisible noise" to "visible semantic rewriting." Validated against comprehensive baselines, SCAR demonstrates that physiological manifold adherence leads to superior transferability and emergent targeted behavior (90.2\% MI mimicry). Furthermore, by exposing the vulnerabilities of the "Human Firewall," SCAR serves as a critical "Red Teaming" tool. Our defense analysis suggests that a Hybrid Adversarial Training strategy is essential to secure diagnostic systems against the full spectrum of threats, from digital noise to biological mimicry.

\clearpage 

\backmatter

\bmhead{Acknowledgments}
This work was supported by Shenzhen University of Advanced Technology and Peking University Shenzhen Hospital.

\bibliography{sn-bibliography}

\end{document}